\RequirePackage{fix-cm}
\documentclass[twocolumn,epjc3]{svjour3}  
\smartqed  
\RequirePackage{graphicx}
\usepackage{latexsym}
\usepackage{hyperref}
\usepackage{dcolumn}
\usepackage{bm}
\usepackage[export]{adjustbox}
\usepackage{siunitx}
\usepackage[square,comma,numbers,compress]{natbib}
\usepackage[utf8]{inputenc}
\usepackage{tabularx}
\usepackage{amsmath}
\journalname{Eur. Phys. J. C}
\begin{document}

\title{Investigation of the Electrical Conduction Mechanisms in P-type Amorphous Germanium Electrical Contacts for Germanium Detectors in Searching for Rare-Event Physics
}


\author{S. Bhattarai\thanksref{USD}
        \and
         R. Panth\thanksref{USD}
         \and
         W.-Z. Wei\thanksref{USD}
         \and
         H. Mei\thanksref{USD} 
         \and
         D.-M. Mei\thanksref{e1,USD}
         \and
         M.-S. Raut\thanksref{USD}
         \and 
         P. Acharya\thanksref{USD}
         \and 
         G.-J. Wang\thanksref{USD}
         }
\thankstext{e1}{dongming.mei@usd.edu}

\thankstext{t1}{ This work was supported in part by NSF NSF OISE 1743790, NSF PHYS 1902577, NSF OIA 1738695, DOE grant DE-FG02-10ER46709, DE-SC0004768, the Office of Research at the University of South Dakota and a research center supported by the State of South Dakota.}


\institute{Department of Physics, University of South Dakota, 414 E Clark St, Vermillion, South Dakota 57069  \label{USD}
}

\date{Received: date / Accepted: date}

\maketitle

\begin{abstract}
For the first time, electrical conduction mechanisms in the disordered material system is experimentally studied for p-type amorphous germanium (a-Ge) used for high-purity Ge detector contacts. The localization length and the hopping parameters in a-Ge are determined using the surface leakage current measured from three high-purity planar Ge detectors. The temperature dependent hopping distance and hopping energy are obtained for a-Ge fabricated as the electrical contact materials for high-purity Ge planar detectors.  As a result, we find that the hopping energy in a-Ge increases as temperature increases while the hopping distance in a-Ge decreases as temperature increases. The localization length of a-Ge is on the order of $2.13^{-0.05}_{+0.07}A^\circ$ to $5.07^{-0.83}_{+2.58}A^\circ$, depending on the density of states near the Fermi energy level within bandgap. Using these parameters, we predict that the surface leakage current from a Ge detector with a-Ge contacts can be much smaller than one yocto amp (yA) at helium temperature, suitable for rare-event physics searches.  
\end{abstract}

\section{Introduction}
\label{intro}
\label{sec:1}
The nature of dark matter and the properties of neutrinos are the important questions of physics beyond the Standard Model of particle physics and remains elusive. Thus, understanding their properties has become an important aspect of underground physics. Numerous research groups are trying to understand their properties by various detection techniques and detection materials~\cite{agostini2019probing,abgrall2018processing,agnese2018first, aalseth2018search, agnese2018nuclear, armengaud2018searches, 2018npa..confE...5K, angloher2009commissioning, yang2019search}. Interaction between dark matter and ordinary matter as a target occurs only through a weakly elastic scattering process, which leaves a very small energy deposition from nuclear or electronic recoils~\cite{mei}. This requires detectors to have a very low-energy threshold. Germanium (Ge) detectors are excellent in the search for dark matter~\cite{armengaud2018searches, agnese2019search,ke2013cdex,aalseth2013cogent}, since Ge detectors offer the lowest energy threshold among the current detector technologies. Also, due to its excellent energy resolution and ability to minimize the background from two neutrino double-beta (2$\nu \beta \beta $) decay, Ge detectors are highly preferred for observing neutrinoless double-beta (0$\nu\beta\beta$) decay~\cite{agostini2015production}. Hence, the high-purity Ge (HPGe) crystals are widely used as detectors for rare event physics. Many research groups like M{\sc ajorana}~\cite{aalseth2018search},
GERDA~\cite{agostini2019probing}, SuperCDMS~\cite{agnese2018first}, CoGeNT~\cite{aalseth2013cogent}, CDEX~\cite{yang2019search} and EDELWEISS~\cite{armengaud2018searches} are using HPGe detectors to detect dark matters and 0$\nu\beta\beta$ decay. A new collaboration named LEGEND~\cite{2018npa..confE...5K} will use tonne-scale $^{76}$Ge detectors in an ultra-low background environment to detect 0$\nu\beta\beta$ decay. These reasons make the fabrication of Ge detectors from HPGe crystals and exploration of their properties an important part of underground physics. A group at the University of South Dakota (USD) has been working on HPGe crystal growth and detector development in order to improve the performance of Ge detectors for rare-event physics searches~\cite{2009APh....31..417M,wang2012development,wang2013optical,mei2015development,wang2014dislocation,yang2015zone,wang2015crystal,wang2018electrical,mei2019impact,wei2018investigation}.
\begin{figure}
    \centering
    \includegraphics[width=\linewidth]{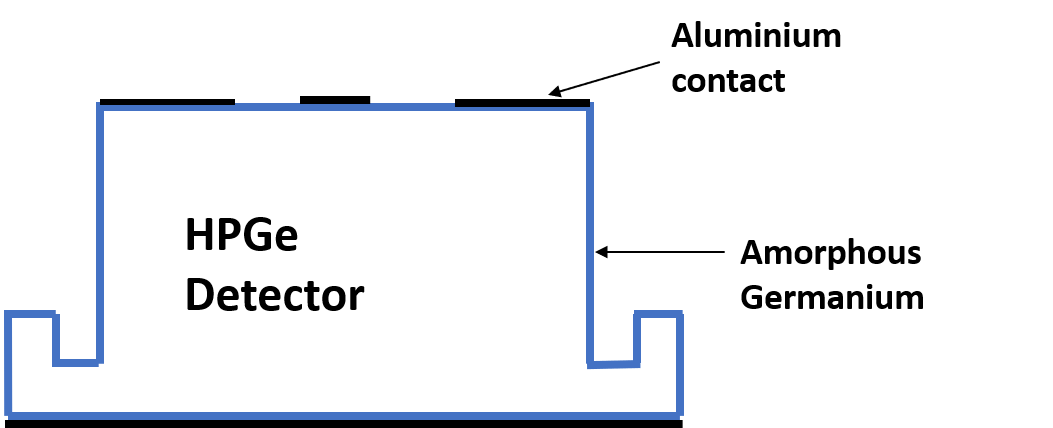}
    \caption{Shown is a schematic diagram of a HPGe planar detector.}
    \label{1}
\end{figure}

A HPGe crystal is fabricated into a planar detector, which is then reversely biased so that it is fully depleted allowing free charge carriers to move. The depletion region acts as an active volume for incident radiation. The energy deposition of incident radiation can be measured by analyzing the interactions in the detector volume~\cite{baertsch1970gamma,llacer1972planar}. The exposed surface of a Ge crystal is sensitive to contamination. The contaminants deposited on the exposed crystal surface can change the electric field distribution in the detector volume that is in close proximity to the exposed surface and  cause a reduction of the resistivity of the surface and hence increase in the surface leakage current. Therefore, a passivation layer is usually applied to protect the exposed surface. This layer should be thin to avoid a large dead layer and it should have large resistivity to prevent excessive leakage current~\cite{baertsch1974surface,tavendale1967semiconductor}. Amorphous Ge (a-Ge)~\cite{hull2005amorphous} and amorphous silicon (a-Si)~\cite{walton1984si} are the most used and accepted passivation layers for semiconductor detectors.
 
\begin{figure}
    \centering
    \includegraphics[width=\linewidth]{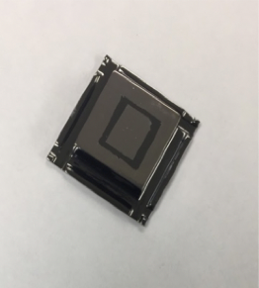}
    \caption{Shown is a Ge detector with a guard ring structure.}
    \label{2}
\end{figure}

 A planar Ge detector fabricated at USD is sketched in Figure~\ref{1}. 
 It consists of a HPGe crystal passivated with a-Ge on the outer surface. The aluminum contact at the bottom is used to provide high voltage. The aluminum contacts on the top are designated for the measurements of the electrical signal including leakage current. The sources of leakage current are: (1) the  bulk leakage current, I$_{bulk}$, which passes through the interior of the detector due to the injection of charge carriers from the contacts and the thermal generation of electron-hole pairs inside the detector volume;  and (2) the surface leakage current, I$_{S}$, which flows through the outer surface of the detector caused by inter-contact surface channels or carrier generation sites. While the bulk leakage current from the USD-fabricated detectors is discussed in detail by Wei et al.~\cite{wenzhao}, the surface leakage current can be misread as the signal which can degrade the performance of the detector. A detector with a guard-ring structure can be used to separate the surface leakage current from the bulk leakage current, allowing us to study the electrical conduction mechanisms in the a-Ge contacts, as shown in Figure~\ref{2}. The passivation material should have high sheet resistivity on the order of greater than  10$^{9}$ ohm/square~\cite{luke2009proximity} to minimize the
 current flowing through the surface. 
 However, even a small amount of current flow through the side surface of the detector can decrease the performance of the detector significantly. Efforts to reduce the surface leakage current require an understanding of the sources of the surface leakage current, which depends upon the electrical properties of the passivating material - a-Ge. Hence, studying the electrical property of a-Ge is crucial for making better passivating materials and reducing the surface leakage current for Ge detectors. 

\begin{figure}
    \centering
    \includegraphics[width=\linewidth]{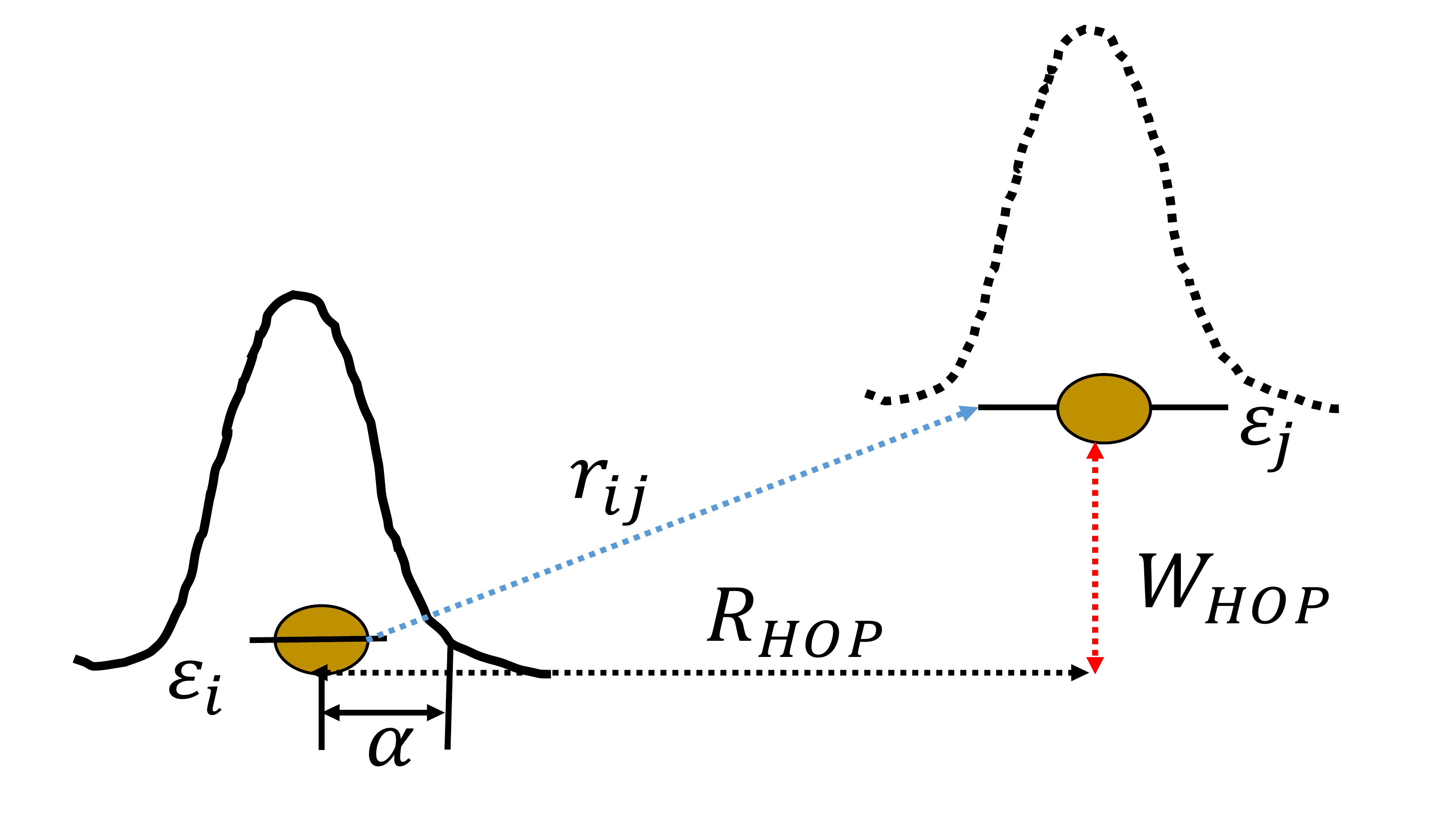}
    \caption{Hopping transition between two localized states $i$ and $j$ with energies of $\varepsilon_i$ and $\varepsilon_j$ , respectively. The solid and dashed lines depict the carrier wavefunctions at sites $i$ and $j$, respectively; $\alpha$ is the localization length; $R_{HOP}$ is the hopping distance; $W_{HOP}$ is the hopping energy. }
    \label{hopping}
\end{figure}

 High resistivity is one of the main requirements for passivating material used in HPGe detectors~\cite{amman2018optimization}. To create a-Ge with high resistivity, hydrogen (7\%) is mixed with argon gas (93\%) to form plasma ions that bombard the Ge target through a sputtering process during detector fabrication. The a-Ge created this way lacks the long-range crystalline order of Ge crystal. Despite having a disordered atomic arrangement, the main features of the electronic band structure are retained in the amorphous phase, including a bandgap quite comparable to the crystalline counterpart. Covalent a-Ge is commonly believed to have localized electronic states at the top of the valence band and the bottom of the conduction band. Unlike in crystalline Ge, the bandgap in a-Ge is occupied by a large number of defect states. Electrical conductivity of a-Ge is thought to be dictated by the hopping mechanism through localized defect states~\cite{mott1969conduction}. Figure~\ref{hopping} depicts an electron from a localized state $i$ to a localized state $j$ that is lower in energy.
 In this localized band, electrons cannot freely travel in space without exchanging some energy with the surrounding environment, usually with phonons, and jump from one state to another. Therefore, this type of conduction is strongly dependent on the density of defects near the Fermi level and the temperature of the material. Since the Ge detectors fabricated with a-Ge contacts are used in liquid nitrogen temperature, we are interested in knowing the properties of a-Ge at low temperatures. Generally, the conduction at low temperature in a-Ge occurs via variable range hopping between localized defect states near the Fermi level. Sir Nevill Mott was one of the first to give a theoretical description of low temperature hopping conductivity in strongly disordered systems~\cite{mott1969conduction,mott1979electron}. In 1969 he introduced the concept of Variable Range Hopping to describe how the long jumps govern the conductivity at sufficiently low temperatures. The electrical conductivity ($\sigma$) of amorphous semiconductors at low temperature (T) obeys the Mott’s relation 
 \begin{equation}
 \label{eqn1}
\sigma=\sigma_{0}e^{-{(T_{0}/T)^{1/4}}}, 
 \end{equation}
 where $\sigma_{0}$ is the conductivity prefactor and T$_{0}$ is the characteristic temperature given by
 \begin{equation}
 \label{eqn2}
     T_{0}=16\alpha^3/kN(\epsilon_f), 
 \end{equation}
 where $\alpha$ is the inverse of localization length and $N(\epsilon_f)$ is the density of defect states near the Fermi level and $k$ is the Boltzmann constant. If we take log of both sides of equation ~\ref{eqn1} and plot the log of conductivity on the y-axis and $T^{-1/4}$ on the x-axis, then we obtain a straight line, the slope of which gives the value of the characteristic temperature $T_0$ and the y-intercept gives the prefactor $\sigma_0$. 
 
 The energy between two localized states (hopping energy) at temperature $T$ is given by
 \begin{equation}
 \label{eqn3}
     W_{HOP}=1/4kT(T_0/T)^{1/4},
 \end{equation}
 and the spatial distance between two hopping sites at temperature $T$ (hopping distance) is
 \begin{equation}
 \label{eqn4}
     R_{HOP}=3/8(T_0/T)^{1/4}\times1/\alpha.
 \end{equation}
 In the past decades, several methods have been used to find the value of the Mott’s parameter for a-Ge by preparing a thin film on a substrate. Yasuda et al.~\cite{yasuda1977effects} found the value of the localization length to be in the range of $5A^\circ$ to $20A^\circ$ for the samples prepared on a glass substrate by the evaporation method. Tolunay et al.~\cite{eray1990evaluation} also studied the electrical properties of evaporated a-Ge at low temperature and found the value of the localization length to be in the range of $8A^\circ$ to $16A^\circ$ using different models compared with the method used by Yasuda et al.~\cite{yasuda1977effects}. Both experiments were performed by preparing the thin films by the evaporation method and the measurements were conducted on pure  a-Ge. In fabricating amorphous contacts on the planar Ge detectors at USD, we use the sputtering method to create a thin film  of a-Ge on the Ge detectors for our study.  Our a-Ge contains a mixture of hydrogen and argon.  Shrestha~\cite{shrestha2011temperature,Shrestha2014Auth} studied the electrical properties of a-Si with different compositions of hydrogen mixtures. The localization length was found to be in the range of $2.13A^\circ$ to $5.07A^\circ$ for different compositions of hydrogen in a-Si. However, there is no report on the evaluation of the Mott’s parameter for the a-Ge used to passivate HPGe detectors. 
 
 In general, the Mott's parameter for a-Ge should be determined through the standard experimental procedure by coating the a-Ge layer onto the surface of an isolating material such a glass substrate. However, one would also like to know the electrical properties of a-Ge coated on the surface of Ge detectors using an well-established fabrication procedure. The goal of this work is to understand the impact of the fabrication procedure on the electrical properties of a-Ge. The variation of the electrical properties between three detectors will provide a range of the surface leakage current for the fabrication procedure and allow us to evaluate if this fabrication procedure can deliver a negligible surface leakage needed for detecting single electron-hole pair at cryogenic temperature. 
 
 We have obtained the values of the localization length $(1/\alpha$), the hopping energy, and the hopping distance of a-Ge for three detectors fabricated at USD. The purpose of this study is to characterize the a-Ge thin layer we created to passivate Ge detectors by comparing our results with the previous work done on similar materials. With such a characterization,  we can revisit our fabrication process to improve the quality of the passivated material and reduce human error, thereby improving the detector performance.
\section {Experimental procedure}
Three HPGe detectors with guard structure, as shown in Figure~\ref{2}, were fabricated with p-type a-Ge passivation in order to study the electrical properties of a-Ge.  Since the planar detector is easier to be fabricated than other geometries and large-size detectors are not required for our study, all detectors used in this work were fabricated into a planar geometry.
A RF sputtering machine was used to sputter a-Ge on all surfaces of the crystal. The thickness of a-Ge, the gas composition of the sputtering process, the pressure, and the applied power can be changed in the fabrication. In this work, a precisely cut crystal in a planar geometry was placed on the jig and loaded into the chamber of the sputtering machine. The plasma was created in the chamber with a mixture of hydrogen and argon gas (7:93) at a pressure of 14 mTorr. The thickness of the a-Ge deposited on the side surface of the crystal is 556 nm and on the top and bottom surfaces of the detector is 1.2 $\mu$m. Although the same deposition apparatus and the same deposition parameters are used to create the a-Ge layers, it is very difficult to maintain the homogeneity of the recipe for detector fabrication process, for example, the time-dependent surface re-oxidation. This may have led to difference in the conductivity of a-Ge for different detectors. This is a main goal of this work to find out the variation of the electrical properties of a-Ge using three detectors fabricated with the same procedure. Additionally, the quality of crystal used to fabricate these detectors and their net impurity concentration, the density of defects, the time since the fabrication, the storage and the handling of the detectors may also contribute to the differences in the electrical properties of a-Ge coated on the Ge detectors.  

\begin{figure}
    \centering
    \includegraphics[width=\linewidth]{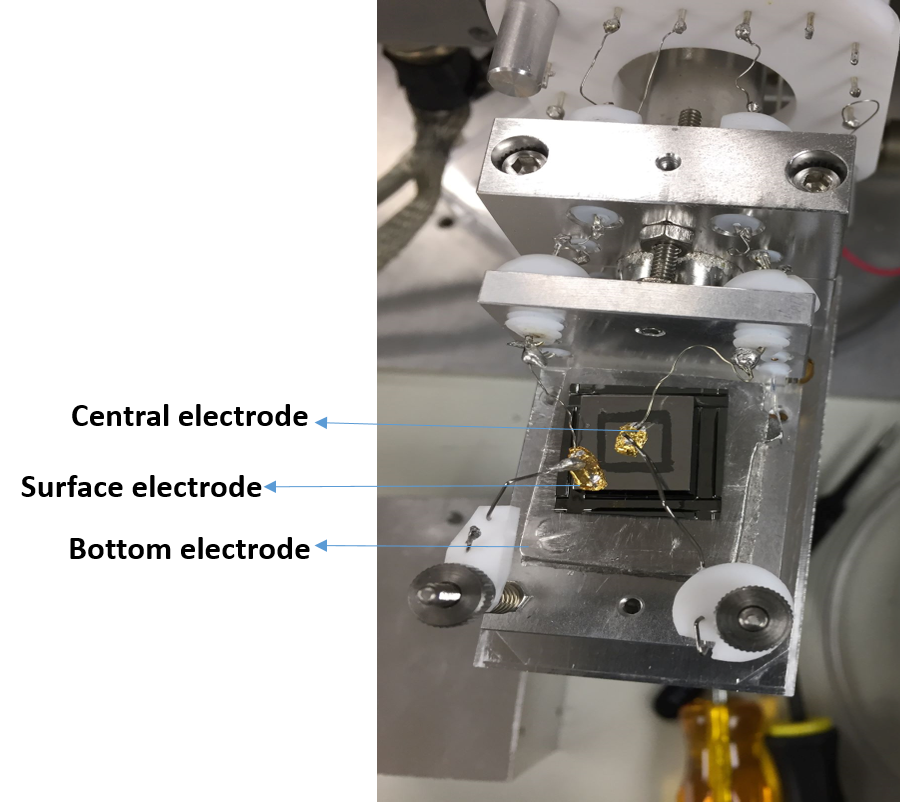}
    \caption{A detector is loaded into a cryostat for I-V measurement at desired temperatures.}
    \label{3}
\end{figure}

After a-Ge was deposited on all surfaces of the crystal, then the detector USD-R02  was loaded into the chamber of an Edwards Electron Beam Evaporator to make the aluminium contacts. An electron beam produced from a tungsten filament bombards the aluminum target. Under high vacuum, the electron beam can reach the crucible without interference. A voltage of 4.89 kV and a current around 35 mA were provided to have a stable data rate of 0.2 to 0.3 nm/s. Note that for the detectors USD-W03 and USD-R03, the aluminium deposition was carried out by sputtering process. The plasma was created in the chamber with argon gas at a pressure of 3 mTorr. A typical thickness for the aluminum contacts was 100 nm. The details are described in an earlier publication from our group~\cite{meng}.
Only the top and bottom surfaces need aluminum contacts to test the electrical properties of a detector. To remove aluminium contacts from the sides, a mask of acid-resisted tape was placed on the top and bottom. Then, the detector was dipped into the acid solution with one percent of HF for a few minutes, until all of the aluminum was etched away from the sides. Note that HF does not remove the a-Ge layer beneath the aluminium.To characterize the electrical properties of a detector, the Ge crystal was loaded into the cryostat, as depicted in Figure~\ref{3}. After the pressure reaches the order of $10^{-6}$ mBar, LN$_{2}$ was added into the Dewar. The temperature of the detector was controlled by the Lakeshore temperature controller. The detector was started at a bias around 50 V and was biased up to 2500 V. The bias voltage was provided to the bottom contact of the detector and the signal was read out from the top contacts. Current-voltage (I-V) characteristic of the surface current for all three detectors was performed by using a transimpedance amplifier, which converts current into voltage. The voltage is then measured by a precision voltmeter. This voltage was then converted back to current, as described in a recent paper from our group~\cite{wenzhao}. The I-V characteristic of two detectors (USD-R03 and USD-WO3) was done at three different temperatures 79K, 90K and 100K, while the I-V characteristic of the detector USD-RO2 was done at 85K, 90K, 95K and 100K.

  \section{\label{sec:level2}Result and Discussion}
\begin{figure}
    \centering
    \includegraphics[width=\linewidth]{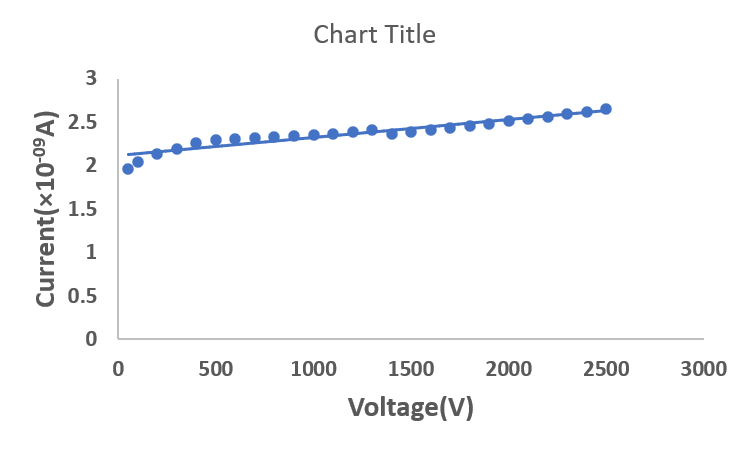}
    \caption{The surface leakage current ($I$) versus voltage ($V$) for USD-W03 at 95K. The reciprocal of the slope of this line gives the resistance at 95K. }
    \label{10}
\end{figure}

Utilizing the first order approximation, the reciprocal of the slope of the I-V curve measured at different temperatures gives the resistance ($R$) of the a-Ge contact layer. As an example, Figure~\ref{10} shows the surface leakage current versus the applied bias voltage for USD-W03 detector. Using this method, we obtained the values of the resistance corresponding to the measured temperatures for three detectors and the results are shown in Table~\ref{tab:my_labels}. 
The resistivity ($\rho$) for a layer of a-Ge with a thickness $t$ on a detector, with a length of sidewall $l$ and a width $w$, was calculated using Ohm's law:
\begin{equation}
\label{eq100}
\rho=4Rtw/l+4Rtw^{'}/l^{'},
\end{equation}
where the constant 4 incorporates the four-side walls of the planar detector, $w^{'}$ represents the width of the wing on the bottom surface of detector and $l^{'}$ is the total length of the groove along which the current flows. A small distance on the top surface from the guard ring to the side surface which contains aluminium was neglected in this study because the resistivity of aluminium is much less than that of a-Ge.
The thickness and the width for USD-R02 are 0.65 cm and 1.4 cm, respectively. For USD-R03, the thickness and the width are 1.6 cm and 0.81 cm.  For USD-W03, the thickness and the width are 0.94 cm and 1.16 cm. For all detectors the value of $t$ is 556 nm, $w^{'}$ is 2 mm and $l^{'}$ is 4.5 mm. Apart from the surface leakage current, the leakage current from the bulk of the detector is also contributed to the surface channel of the detector. This current should be subtracted from the surface leakage current in order to study the electrical properties of a-Ge. A theoretical model that describes the current voltage relationship for amorphous-crystalline heterojunction was developed by D$\Ddot{o}$hler and Brodsky~\cite{sze1981physics,henisch1958rectifying}. For a-Ge coated on the surface of Ge, the energy barrier height for hole and electron injections are represented by $\phi_h$ and $\phi_e$, respectively; the effective Richardson constant is $A$, the barrier lowering terms are $\Delta\phi_h$  and $\Delta\phi_e$,  which account for the lowering of hole and electron energy barrier height, respectively due to the penetration of the electric field into the a-Ge contacts. Putting all of these parameters in an equation,  the current density $J$ is given by~\cite{sze1981physics,henisch1958rectifying}
\begin{equation}
\begin{split}
   J & = A^*T^2exp[-(\phi_h-\Delta\phi_h/kT)],\\                          \text{ where } & \Delta\phi_h=\sqrt{2qV_a N_d/N_f},
   \label{eqn116}
   \end{split}
\end{equation} and
\begin{equation}
    \begin{split}
   J & = A^*T^2exp[-(\phi_e-\Delta\phi_e/kT)], \\                          \text{ where } & \Delta\phi_h=\sqrt{\epsilon_0\epsilon_{Ge}/N_f}(V_a-V_d)/t.
   \end{split}
   \label{eqn117}
\end{equation}
Note that equations ~\ref{eqn116} and ~\ref{eqn117} represent the current density before and after the full depletion of the detector, respectively. $ N_d$ is the net ionized impurity concentration of the detector, $N_f$ is the density of localized energy states (defects) near the Fermi level in a-Ge, $k$ is the Boltzmann constant, $\epsilon_0$ is the free-space permittivity, $\epsilon_{Ge}$ is the relative permittivity for Ge, $V_d$ is the full depletion voltage and $t$ is the detector thickness, $q$ is the magnitude of the electron charge, $V_a$ is the applied biased voltage. The sum of equations ~\ref{eqn116} and ~\ref{eqn117} give the total current density after the detector is fully depleted. The current injected into the bulk from the contacts was calculated by using the area of the aluminium contact outside the guard ring. These areas for USD-R02, USD-R03 and USD-W03 were 1.79 $cm^2$, 1.84 $cm^2$ and 0.98 $cm^2$ respectively. The values of $\Delta\phi_h$ , $\phi_h$ and $N_f$ have been calculated for these detectors in our group ~\cite{wenzhao}.  \\

The results for the calculated conductivity are shown in Table~\ref{tab:my_labelss}.
\begin{table}
\centering
\begin{tabular}{|c|c|c|c|}
    \hline
    &\multicolumn{3}{|c|}{Detector's Resistance($\SI{}{\ohm}$)}\\
    \hline
    Temperature&{USD-R03}&{USD-R02}&{USD-W03}\\
    \hline
    79&$2\times10^{14}$&-&$5\times10^{14}$\\
    \hline
    85&-&$1.1\times10^{14}$&-\\
    \hline
    90&$2\times10^{13}$&$2.5\times10^{13}$&$1.4\times10^{13}$\\
    \hline
    95&$2.5\times10^{12}$&$1\times10^{13}$&$5\times10^{12}$\\
    \hline
    100&-&$5\times10^{12}$&-\\
    \hline
    \end{tabular}

\caption{The calculated values of the resistance from the I-V curves for three USD fabricated detectors.}
\label{tab:my_labels}
\end{table}

\begin{table*}
\centering

\begin{tabular}{|c|c|c|c|c|c|c|c|c|c|c|}
\hline
Detector&\multicolumn{3}{|c|}{USD-R03}&\multicolumn{4}{|c|}{USD-R02}&\multicolumn{3}{|c|}{USD-W03}\\
\hline
Temperature (K)&79&90&95&85&90&95&100&79&90&95\\
\hline
Conductivity($\SI{}{10^{-12}\ohm}^{-1}cm^{-1}$)&3.5&35.0&280.1&6.1&27.1&68.1&130.6&1.5&55.1&157.3\\
\hline

\end{tabular}
\caption{The calculated values of the conductivity ($\sigma)$ for three USD-fabricated detectors.}
\label{tab:my_labelss}
\end{table*}

\begin{figure}
\centering
  \includegraphics[width=\linewidth]{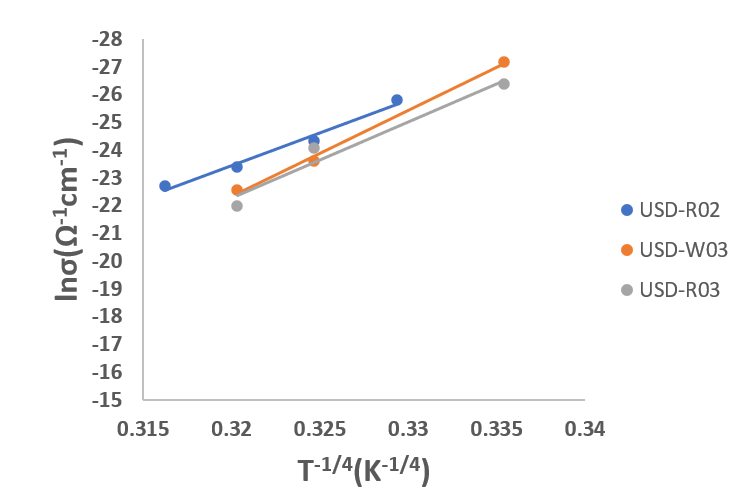}
     \caption{The variation of conductivity with temperature for detectors USD-R02, USD-R03 and USD-W03. The slope of the plot for USD-R02 is found to be -234.2 and the Y-intercept is 51.5. Similarly, the slope for USD-R03 is found to be -275.6 and the Y-intercept is 65.9. Likewise, the slope and the Y-intercept for USD-W03 are found to be -309.6 and 76.7, respectively.}
    \label{4}
\end{figure}

The variation of conductivity with temperature is studied for three different detectors, as shown in Figure ~\ref{4}.  The slopes of the fitted straight lines are used to calculate the characteristic temperature ($T_{0}$) and the intercepts are used to obtain the conductivity prefactor ($\sigma_{0}$) for three a-Ge layers used as the contacts for three Ge detectors.  The electrical conductivity of the a-Ge sputtered on a HPGe detector in the low temperature range was studied by Amman et al.~\cite{amman2018optimization}. The a-Ge contacts fabricated in this work was performed using a similar recipes (7$\% $ Hydrogen, 11 mTorr pressure). There is a significant variation of conductivity of a-Ge measured in this work with the similar work done by Amman et al. In the referred work a-Ge was sputtered on a glass substrate and the pressure used to sputter was 11 mTorr. We used 14 mTorr pressure with same hydrogen argon composition ratio and the substrate we used was a HPGe crystal. The differences in the conductivity can affect the values of the Mott’s Parameter. Therefore, the Mott's parameters should be determined for a-Ge fabricated with a specific machine. The three detectors used in this study show similar ranges of conductivity. Thus, the values of the localization length, the hopping energy and the hopping distance reported in this work are for the USD fabricated detectors. Table~\ref{tab:T0} shows the calculated characteristic temperature ($T_0$) and the conductivity prefactor ($\sigma_0$) for three USD-fabricated detectors. Although the a-Ge layers in three detectors have similar thickness, the measured values of the density of defects $N_f$ and the barrier heights $\phi_h$ and $\phi_e$ are different ~\cite{wenzhao}. Also the net impurity concentration for all the detectors is different so that the barrier lowering term $\Delta\phi_h$ and $\Delta\phi_e$ for the a-Ge layers are different. The fabrication handling process and the time of storage of these detectors are also different. These factors may have contributions to the time-dependent surface re-oxidation, which contributes to the difference in the measured properties of the a-Ge coated on the surface of Ge detectors. 
\begin{table}
\centering
\label{tab:T0}

\begin{tabular}{|c||c||c|}
\hline
Detector&$T_0$(K)&$\sigma_0$($\SI{}{\ohm}^{-1}cm^{-1}$)\\
\hline
USD-R02&$3.04\times10^{9}$&$2.30\times10^{22}$\\
\hline
USD-W03&$9.19\times10^{9}$&$2.03\times10^{33}$\\
\hline
USD-R03&$5.77\times10^{9}$&$4.16\times10^{28}$\\
\hline

\end{tabular}

\caption{The calculated values of characteristic temperature ($T_0$) and conductivity prefactor ($\sigma_0$) for three USD-fabricated detectors.}
\label{tab:my_labe}
\end{table}
\begin {table*}
\scalebox{0.9}{%
 \begin{tabular}{|c|c|c|c|c|c|c|c|c|c|c|c|c|}
    \hline\
    Detector&\multicolumn{3}{|c|}{USD-R03}&\multicolumn{3}{|c|}{USD-R02}&\multicolumn{3}{|c|}{USD-W03}\\
    \hline
    Temperature&$1/\alpha$$(A^\circ)$&$W_{HOP}$$(meV)$&$R_{HOP}$$(A^\circ$)&$1/\alpha$$(A^\circ$)&$W_{HOP}$$(meV)$&$R_{HOP}$$(A^\circ$)&$1/\alpha$$(A^\circ$)&W$_{HOP}$$(meV)$&R$_{HOP}$$(A^\circ$)\\
    \hline
79&	$2.2^{-0.26}_{+0.58}$&	157.2&75.9&	$5.07^{-0.83}_{+2.58}$&	-&	-&	$2.13^{-0.05}_{+0.07}$&	176.6&	82.5\\
\hline
85&	$2.2^{-0.26}_{+0.58}$&	-&	-&$5.07^{-0.83}_{+2.58}$	&	141.5&147.0&	$2.13^{-0.05}_{+0.07}$&	-&	-\\
\hline
90&$2.2^{-0.26}_{+0.58}$	&	173.3&	73.5&	$5.07^{-0.83}_{+2.58}$&	147.7&	144.9&	$2.13^{-0.05}_{+0.07}$&	194.7&	80.0\\
\hline
95&	$2.2^{-0.26}_{+0.58}$&	180.5&	72.5&	$5.07^{-0.83}_{+2.58}$&	153.8&	143.0&	$2.13^{-0.05}_{+0.07}$&	202.8&	78.8\\
\hline
100&	$2.2^{-0.26}_{+0.58}$&	-&	-&	$5.07^{-0.83}_{+2.58}$&	159.8&	141.1&	$2.13^{-0.05}_{+0.07}$&	-&	-\\
\hline
    \end{tabular}}
    \caption{The measured values of the localization length, the hopping energy and the hopping distance for three USD detectors.}
    \label{tab:my_label}
\end{table*}

The value of the characteristic temperature T$_0$ is calculated for each detector from the slope of these plots in Figure~\ref{4}. The variation of T$_{0}$ reflects the difference in the density of states near the Fermi level for three different a-Ge layers. The values of the density of states near the Fermi level $N(\epsilon_f)$ for these detectors are obtained in a recent paper from our group~\cite{wenzhao}. The  value for USD-R02 is found to be N($\epsilon_f)$= ($4.68\pm3.32)\times10^{17}eV/cm^3$. Since there are two values of N($\epsilon_f)$ corresponding to two contacts for USD-R03, we simply take the average of these two values to obtain the density of states for USD-R03 and the average value used to calculate the Mott's parameter in this study is $N(\epsilon_f)$= $3.08^{+1.36}_{-1.58}\times10^{18}eV/cm^3$. Similarly, the average value of density of states for USD-W03  is found to be $N(\epsilon_f)$= $2.1^{+0.17}_{-0.20}\times10^{18}eV/cm^3$. With these values of $N(\epsilon_f)$ and T$_0$ determined and the Boltzmann constant $k$,  the value of $\alpha$ can be calculated using the equation~\ref{eqn2}. The calculated values of the localization length for detector USD-R02, USD-R03 and USD-W03 are   $5.07^{-0.83}_{+2.58}A^\circ$, $2.2^{-0.26}_{+0.58}
A^\circ$, and $2.13^{-0.05}_{+0.07}A^\circ$, respectively. Table~\ref{tab:my_label} displays the results obtained in this work. The errors are dictated by the errors from the density of states near the Fermi level. 

The values of the localization length obtained for the a-Ge fabricated at USD are less than the values reported previously~\cite{szpilka1982dc,eray1990evaluation,yasuda1977effects}.
This difference in localization length can be attributed to the difference in the fabrication of a-Ge between the previous work and our work. The previous work referenced in this work used pure a-Ge, while we used hydrogenated a-Ge. A similar work on hydrogenated a-Si was reported and their results are comparable to our work~\cite{Shrestha2014Auth}. The value of the localization length is directly related to the density of defects $N(\epsilon_f)$  and $T_0$. The amount of hydrogen reduces the density of defect states significantly and hence increases the resistivity of a-Ge. This suggests that a-Ge can be fabricated with or without hydrogen content, depending on the applications. If high resistivity is preferred, such as the passivation for Ge detectors, the a-Ge should be fabricated with hydrogen content. If low resistivity is needed, such as solar cells, the a-Ge should be made without hydrogen content. This is to say that if the recipe of a-Ge deposition is modified, then the film's resistivity~\cite{amman2018optimization,Looker2014Auth} and hence the Mott’s parameter are impacted by the fabrication process. Because we determine the electrical property of hydrogenated a-Ge passivated on HPGe detectors deposited by sputtering method and the referenced work considers pure a-Ge on a thin films on a substrate by the evaporation method,  the difference in the localization length can be expected. However, all the calculated values of the localization length are in the acceptable range~\cite{yasuda1977effects,eray1990evaluation,Shrestha2014Auth,szpilka1982dc}.

\begin{figure}
    
\includegraphics[width=\linewidth]{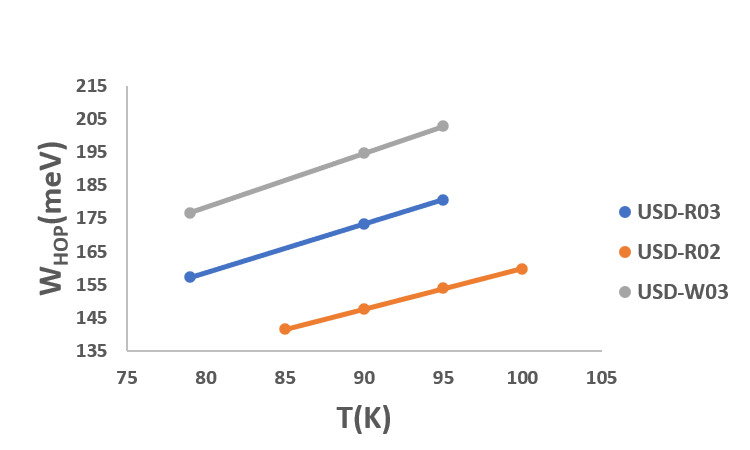}
\caption{Shown is the variation of hopping energy with temperature for three different detectors.}
\label{5}
\end{figure}

In addition, the hopping energy and the hopping distance are calculated for each of the detectors using equations~\ref{eqn3} and \ref{eqn4}, respectively. The variation of hopping energy W$_{HOP}$ with temperature (T) is also studied for all three detectors, as shown in Figure~\ref{5}.\\

\begin{figure}
    \centering
    \includegraphics[width=\linewidth]{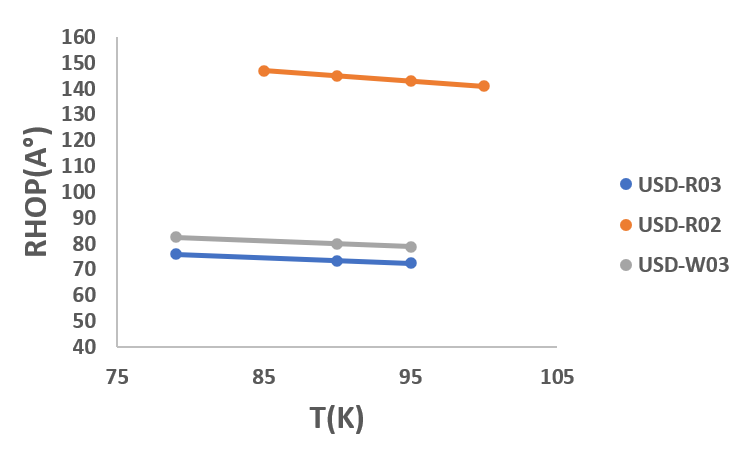}
    \caption{The variation of hopping length versus temperature for three different detectors.}
    \label{6}
\end{figure}

The value of hopping energy increases with the increase in temperature. We obtain a larger value of T$_0$  as compared with a similar work for the  a-Ge made without hydrogen. This indicates that the value of hopping energy is larger in our a-Ge. A larger hopping energy means that the charge carriers jumping from one defect state to another defect state for conduction require higher kinetic energy, which make the conduction process difficult and hence the material is highly resistive.
Similarly, the variation of hopping length R$_{HOP}$ with temperature (T) is also studied as shown in Figure~\ref{6}.

\begin{figure}
\centering
\includegraphics[width=\linewidth]{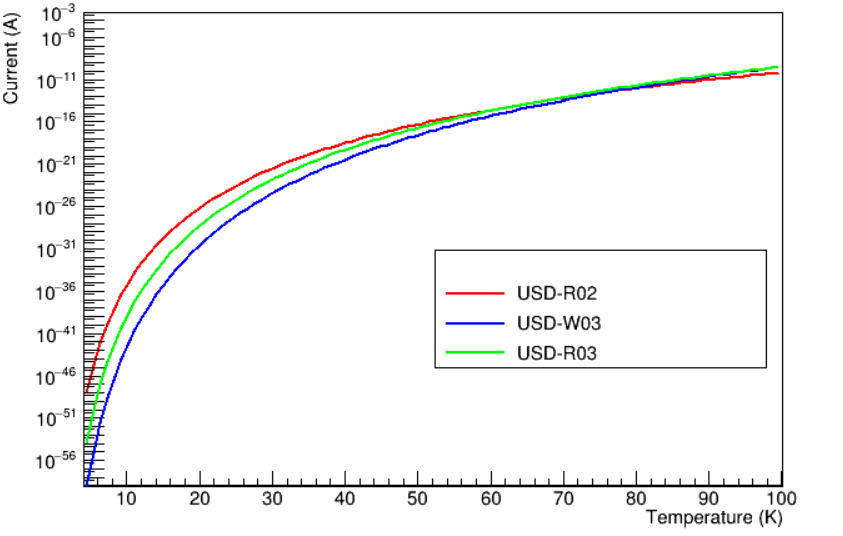}
\caption{Projected variation of the surface leakage current with temperature for a PPC detector using the parameters obtained for the a-Ge used in detectors USD-R03, USD-R02 and USD-WO3.}
\label{7}
\end{figure}
\begin{figure}
\centering
\includegraphics[width=\linewidth]{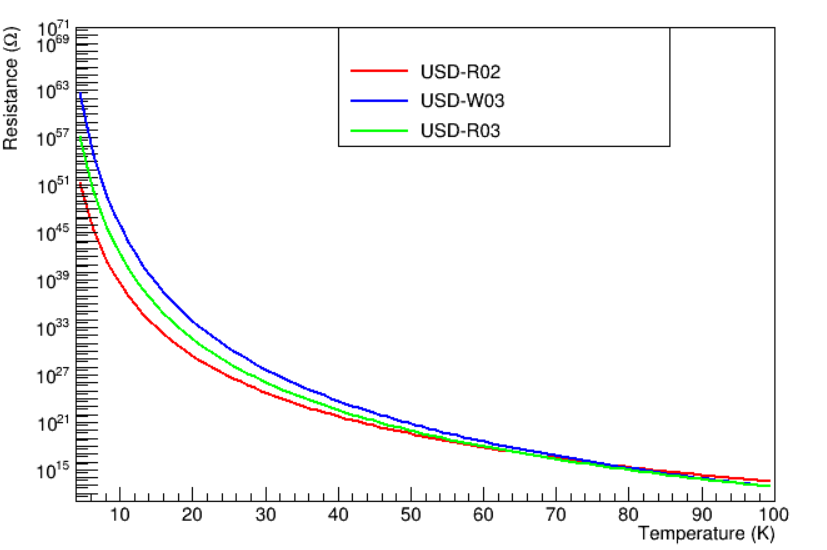}
\caption{Projected variation of the resistance with temperature for a PPC detector using the parameters obtained for the a-Ge used in detectors USD-R03, USD-R02 and USD-WO3.}
\label{100}
\end{figure}

From this study we find that the hopping length R$_{HOP}$ decreases with increasing in temperature. R$_{HOP}$, as indicated in equation \ref{eqn4},  is small for small values of localization length.  Thus, the wave function is more localized for trapping charges, making it difficult for them to hop to other trap states, resulting in the increase of resistance and hence the resistivity. The calculated values of R$_{HOP}$ and the localization length ($1/\alpha$) are lower than the similar work reported previously without hydrogen content. This suggests that the a-Ge created with hydrogen has higher resistance and resistivity, suitable for passiviting Ge crystals when making Ge detectors.

Using the values of $T_0$ and $\sigma_0$,  we can estimate the amount of the surface leakage current in a HPGe detector with a-Ge contact at helium temperature, assuming both $T_0$ and $\sigma_0$ are temperature independent. For an a-Ge passivation of the thickness $t$ on a HPGe P-type point contact (PPC) detector of a length $l$ and a radius $r$, the resistance $R$ is given by
 \begin{equation}
     R=l/(\sigma\pi t(2r+t)),
 \end{equation}
 where $\pi t(2r+t)$ gives the cross-sectional area of the annular portion of a-Ge on the detector. By knowing the value of the conductivity $\sigma$ from equation~\ref{eqn1}, we can find the resistance of the a-Ge at various temperatures. Thus, for a given bias voltage, $V$, we can estimate the value of the surface leakage current, $I_{S}$, at different temperatures. We are particularly interested in the surface leakage current at very low temperature such as liquid helium temperature.
 
To predict the surface leakage current at liquid helium temperature, we assume a PPC detector of 1.02 kg mass with 7 cm in diameter and 5 cm in length. The thickness of the a-Ge  passivation layer is assumed to be 556 nm. Using the values of $\sigma_0$ and $T_0$ from Table ~\ref{tab:my_labe},  we estimate the surface leakage current and the resistance in the low temperature regime, as shown in Figures~\ref{7} and~\ref{100}. It is clear that the surface leakage current is extremely small (nearly zero) at liquid helium temperature of 4 K.  

\section{\label{sec:level3}CONCLUSION}
We have determined the values of the Mott's parameters for three a-Ge layers used as planar Ge detector contacts fabricated at USD. As a result, we find that the localization length of a-Ge is on the order of $2.13^{-0.05}_{+0.07}A^\circ$ to $5.07^{-0.83}_{+2.58}A^\circ$, depending on the density of states near the Fermi energy level within bandgap. The hopping energy ranges from 141.5 meV to 202.8 meV and the hopping distance varies from 72.5 $A^\circ$ to 147.0 $A^\circ$, depending largely on temperature. We find that the hopping energy in a-Ge increases as temperature increases while the hopping distance in a-Ge decreases as temperature increases.
Our results are different from that of pure a-Ge fabricated without hydrogen content, but comparable to a-Si fabricated with hydrogen content. This study confirms that the amount of hydrogen can reduce the density of defect states near the Fermi level significantly and hence can increase the resistivity of a-Ge. Subsequently, the values of the characteristic temperature T$_0$ and the localization length ( $1/\alpha$ ) obtained in this study indicate a high resistivity of the a-Ge fabricated with hydrogen content at USD. The high resistivity of a-Ge is an essential characteristic of a good passivation material for HPGe detectors. The variation of the hopping energy, the hopping distance, and the localization length in three different a-Ge layers corresponds to the difference in the density of states near the Fermi level, which reflects the variation of the fabrication process for making a-Ge layers. The time-dependent re-oxidation and personal errors in the fabrication process may also have led to difference in these parameters. An in-depth study of the effects of surface re-oxidation is mandatory to reduce the scattering in the measured values in order to achieve the complete control of the production process of HPGe detectors. The values of the parameters calculated in this study shows that the a-Ge fabricated at USD to passivate Ge detectors meet the criteria for passivation. Using the parameters of the localization length, the hopping energy, and the hopping distance, we predict that the surface leakage current for a PPC detector with a-Ge contacts at helium temperature (4 K) is nearly zero, suitable for light dark matter searches.

\section*{Acknowledgments}
The authors would like to thank Mark Amman for his instructions on fabricating planar detectors and Christina Keller for a careful reading of this manuscript. We would also like to thank the Nuclear Science Division at Lawrence Berkeley National Laboratory for providing us a testing cryostat.

%
%


%
%

\bibliographystyle{spphys}       
\bibliography{sanjay.bib}  

%
%

\end{document}